\documentclass[pre,aps,preprint,amsmath]{revtex4}
\usepackage{fancyhdr}
\usepackage{graphicx}
\usepackage{dcolumn}
\usepackage{txfonts}
\usepackage{bm}
\begin{document}

\title{Realizing chameleonlike thermal rotator with transformation-invariant metamaterials}

\author{Fubao Yang}
\email{18110190009@fudan.edu.cn}
\affiliation{Department of Physics, State Key Laboratory of Surface Physics, and Key Laboratory of Micro and Nano Photonic Structures (MOE), Fudan University, Shanghai 200438, China}
\author{Boyan Tian}
\affiliation{Department of Physics, State Key Laboratory of Surface Physics, and Key Laboratory of Micro and Nano Photonic Structures (MOE), Fudan University, Shanghai 200438, China}
\author{Liujun Xu}
\email{13307110076@fudan.edu.cn}
\affiliation{Graduate School of China Academy of Engineering Physics, Beijing 100193, China}

\begin{abstract}
Heat flux rotation has important significance in thermal protection since it can shield the heat energy from a selected direction. Combining with tailored metamaterials, transformation thermotics provides a powerful way to manipulate heat flux, and various kinds of thermal meta-devices have been designed including thermal rotator. However, the existing transformation-thermotics-based thermal rotator can only work in a fixed background. Remanufacturing is inevitable when background changes, which is inconvenient and restricts the practical application. Here, we propose a novel mechanism for chameleonlike thermal rotator. The designed rotator can adaptively change its thermal conductivity with the object nearby while rotating heat flux without distorting the background temperature profile, just like a chameleon in nature. Moreover, such rotator is made of transformation-invariant material, thus its constitutive parameters do not change under arbitrary coordinate transformations. Therefore, the proposed rotator also has functionality-invariance beyond shape adjustment, and can theoretically transfer heat flux in arbitrary direction using different shapes of the same material.  A prototype rotator was designed and fabricated, and its chameleonlike behavior is successfully demonstrated. Our concept provides a guidance to design chameleonlike thermal meta-devices and can be extended to other fields like acoustics, hydrodynamics, etc. The chameleonlike thermal rotator will have potential applications for the implementation of adaptive and adjustable metamaterials.
\end{abstract}

\maketitle
\section{Introduction}
Transformation thermotics~\cite{APL2008,APL2008-1} offers a powerful means to manipulate heat flux at will. The initial investigations were predominantly centered around thermal conduction~\cite{GaoEPL13,ZhuAIP15,LiPLA16,ShenAPL16,YangAPL17,WangJAP18,XuJAP18,XuPRE18,XuPRAP19-1,XuPRAP19-2,XuPRE19,JinIJHMT20,DaiIJHMT20,WangEPL21,LiuJAP21,JinIJHMT21,TianIJHMT21,LiuJAP21-2,HuangAMT22,ZhuangSC22,YangPRAP22,ZhuangPRE22,ZhouEPL23,LeiIJHMT23,YangPRAP23,DaiCSF23}, leading to the proposal of various functionalities such as cloaking, concentration, and rotation~\cite{Huang,Wang,YangPR21}. To cater to practical applications, the exploration extended to considering convection~\cite{AIP2015,PRE2018,DaiJAP2018,ESEE2019,IJHMT2020,CPL2020,XuCPL20,SC2020,XuIJHMT21,XuEPL21-1,WangPRAP21,XuPRE21,XuAPL21,XuEPL21-2,DaiPRAP22,XuPRL22,XuPRL22-2} and radiation~\cite{XuPRAP19,PRAP2020T,ESEE2020,WangPRAP20}, leading to the development of corresponding transformation theories.

In the realm of transformation-thermotics-based metamaterials, significant progress has been achieved. However, the persisting challenge lies in the lack of adaptability. Specifically, the pivotal equation of transformation thermotics, represented as $\tensor\kappa'={\rm\bm{J}}\tensor\kappa{\rm\bm{J}}^\tau/\det{\rm\bm{J}}$, where $\tensor\kappa'$ signifies transformed thermal conductivity, $\tensor\kappa$ represents environmental thermal conductivity, ${\rm\bm{J}}$ stands for the Jacobian matrix, and $\tau$ denotes transpose. Clearly, the transformed parameter $\left(\tensor\kappa'\right)$ is significantly reliant on the environmental parameter $\left(\tensor\kappa\right)$. In other words, any changes in the environment would necessitate corresponding changes in the transformed parameter, causing the original design to falter in new surroundings. This limitation is substantial, as each device would only be functional within a specific environment. Similar constraints are also observed in electromagnetic and thermal radiation fields, with several insightful studies shedding light on this issue~\cite{PRX2017,NC2018}.

\begin{figure}
\centering
\includegraphics[width=1\linewidth]{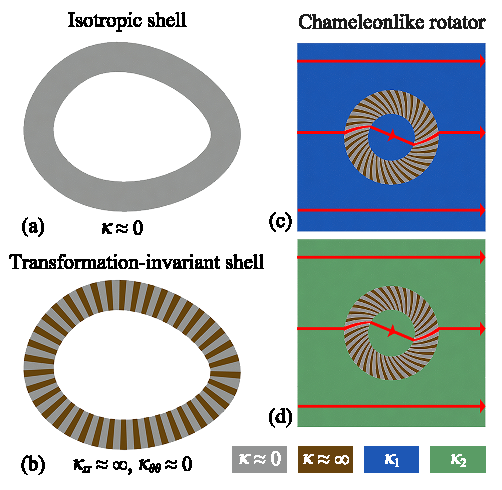}
\caption{\label{1} Schematic diagram of thermal chameleonlike rotator. (a) Isotropic shell with near-zero thermal conductivity. (b) Transformation-invariant shell with near-zero thermal conductivity in tangential direction and near-infinite thermal conductivity in radial direction. (c) and (d) Thermal chameleonlike rotator working in different environments. Lines with arrows indicate heat flow. The environmental thermal conductivities of (c) and (d) are $\kappa_1$ and $\kappa_2$, respectively. Adapted from Ref.~\cite{YangPRAP20}.}
\end{figure}

With the goal of enhancing the adaptability of thermal metamaterials, we propose a mechanism grounded in thermal transformation-invariant metamaterials~\cite{ATS2018,OE2019}, which exhibit highly anisotropic thermal conductivities—0~W~m$^{-1}$ K$^{-1}$ in one direction and $\infty$~W~m$^{-1}$ K$^{-1}$ in the other~\cite{AEM2015,AM2017,IJHMT2018}. Notably, transformation-invariant (i.e., highly anisotropic) metamaterials have elicited widespread interest in diverse fields such as electromagnetism~\cite{JO2018,PRL2019} and acoustics~\cite{PRAP2019W,PRAP2019F}. In a two-dimensional scenario, these highly anisotropic thermal conductivities manifest adaptive responses to environmental shifts~\cite{PRAP2019,SC2020}, akin to chameleons. By implementing coordinate transformations based on transformation-invariant metamaterials, we maintain this chameleon-like behavior, enabling designed devices to adapt to varying environmental conditions. Taking thermal rotators~\cite{PRL2012,PRE2019} as an illustration, which govern the direction of heat flux, current designs are restricted to specific environments and lack flexibility in dealing with environmental changes. To enhance adaptability, we introduce the concept of thermal chameleon-like rotators. Departing from a conventional isotropic shell with nearly zero thermal conductivity [refer to Fig.~\ref{1}(a)], we initiate rotation transformation from a transformation-invariant shell [refer to Fig.~\ref{1}(b)]. As a result, the devised rotator can adeptly function across different environments [refer to Figs.~\ref{1}(c) and \ref{1}(d)], earning the label of a chameleon-like rotator. Here, "environment" pertains to regions excluding the rotator, while the "environmental parameter" refers to thermal conductivity. The scheme's validation is further carried out through simulations and experiments. Let us now delve into the theoretical aspect.

\section{Theoretical methods}
We explore a passive and stable two-dimensional conduction process, which adheres to the Fourier law:
\begin{eqnarray}
	\nabla\cdot\left(-\tensor\kappa\cdot\nabla T\right)=0.\label{E1}
\end{eqnarray}
The entire system divides into three distinct regions: the core, shell, and background, each having tensorial thermal conductivities of $\tensor\kappa_{1}=\kappa_{1}\tensor I$, $\tensor\kappa_{2}=\rm{diag}\left(\kappa_{rr},\kappa_{\theta\theta}\right)$, and $\tensor\kappa_{3}=\kappa_{3}\tensor I$, respectively. The core and background are treated as the environment, assuming equal thermal conductivities $\kappa_1=\kappa_3$. In cylindrical coordinates $\left(r,,\theta\right)$, $\tensor\kappa_{2}$ takes form. The effective thermal conductivities $\kappa_e$ of the core and shell can be obtained by solving the Laplace equation:
\begin{eqnarray}
	\kappa_{e} = \kappa_{rr}\frac{n_1\left(\kappa_{1}-n_{2}\kappa_{rr}\right)-n_2\left(\kappa_{1}-n_{1}\kappa_{rr}\right)p^{\left(n_{1}-n_{2}\right)/2}}{\kappa_{1}-n_{2}\kappa_{rr}-\left(\kappa_{1}-n_{1}\kappa_{rr}\right)p^{\left(n_1-n_2\right)/2}},\label{E2}
\end{eqnarray}
where $n_{1,,2} = \pm \sqrt{\kappa_{\theta\theta}/\kappa_{rr}}$ and $p = \left(R_{1}/R_{2}\right)^{2}$. $R_1$ and $R_2$ represent the inner and outer radii of the shell. The thermal conductivity of transformation-invariant metamaterials appears as:
\begin{eqnarray}
	\tensor\kappa_{2} =
	\begin{pmatrix}
		\infty & 0\\
		0 & 0
	\end{pmatrix}.\label{E3}
\end{eqnarray}
Substituting Eq.~(\ref{E3}) into Eq.~(\ref{E2}), we find that the effective thermal conductivity $\kappa_{e}$ approximates $\kappa_{1}$, implying adaptability to changing environments. This implies that two-dimensional transformation-invariant metamaterials exhibit a chameleon-like behavior.

Subsequently, we delve into an arbitrary two-dimensional coordinate transformation:
\begin{subequations}\label{E6}
	\begin{align}
		r'=R\left(r,,\theta\right),\
		\theta'=\Theta\left(r,\theta\right),
	\end{align}
\end{subequations}
where $\left(r',,\theta'\right)$ denote physical coordinates and $\left(r,,\theta\right)$ represent virtual coordinates. The Jacobian matrix ${\rm\bm{J}}$ takes the form:
\begin{eqnarray}
	{\rm\bm{J}}=
	\begin{pmatrix}
		\frac{\partial r'}{\partial r} & \frac{\partial r'}{r\partial \theta}\\
		\frac{r' \partial \theta'}{\partial r} & \frac{r' \partial \theta'}{r\partial \theta}
	\end{pmatrix}.\label{E7}
\end{eqnarray}
The transformed thermal conductivity becomes:
\begin{eqnarray}
	\tensor\kappa_{2}' = \frac{{\rm\bm{J}}\tensor\kappa_{2}{\rm\bm{J}}^\tau}{\det{\rm\bm{J}}},\label{E9}
\end{eqnarray}
detailed as:
\begin{widetext}
	\begin{eqnarray}
		\tensor\kappa_{2}' = \frac{1}{\det{\rm\bm{J}}}
		\begin{pmatrix}
			\kappa_{rr}\left(\frac{\partial r'}{\partial r}\right)^{2} + \kappa_{\theta\theta}\left(\frac{\partial r'}{r\partial \theta}\right)^{2} & \kappa_{rr}\left(\frac{\partial r'}{\partial r}\right)\left(\frac{r'\partial \theta' }{\partial r}\right) + \kappa_{\theta\theta}\left(\frac{\partial r'}{r\partial \theta}\right)\left(\frac{r'\partial \theta' }{r\partial \theta}\right)\\ \kappa_{rr}\left(\frac{\partial r'}{\partial r}\right)\left(\frac{r'\partial \theta' }{\partial r}\right) + \kappa_{\theta\theta}\left(\frac{\partial r'}{r\partial \theta}\right)\left(\frac{r'\partial \theta' }{r\partial \theta}\right) & \kappa_{rr}\left(\frac{r'\partial \theta' }{\partial r}\right)^{2} + \kappa_{\theta\theta}\left(\frac{r'\partial \theta' }{r\partial \theta}\right)^{2}
		\end{pmatrix}.\label{E10}
	\end{eqnarray}
\end{widetext}
Using Eq.~(\ref{E3}), the eigenvalues of Eq.~(\ref{E10}) are:
\begin{subequations}\label{E11}
	\begin{align}
		\lambda_{1} = \frac{\kappa_{rr}}{\det{\rm\bm{J}}} \left[\left(\frac{\partial r'}{\partial r}\right)^{2}+\left(\frac{r' \partial \theta'}{\partial r}\right)^{2}\right],\\
		\lambda_{2} \approx \frac{\kappa_{\theta\theta}}{\det{\rm\bm{J}}}.
	\end{align}
\end{subequations}
Due to $\kappa_{rr} = \infty $ and $\kappa_{\theta\theta} = 0$, Eq.~(\ref{E11}) simplifies to:
\begin{subequations}\label{E12}
	\begin{align}
		\lambda_{1} = \infty,\
		\lambda_{2} = 0.
	\end{align}
\end{subequations}
Clearly, an arbitrary coordinate transformation does not alter the eigenvalues.

Our next step involves designing a thermal chameleon-like rotator utilizing transformation-invariant metamaterials. The rotational coordinate transformation can be expressed as:
\begin{subequations}\label{E13}
	\begin{align}
		r' = r,\
		\theta' = \theta + \theta_{0}~~~~\left(r < R_{1}\right),\\
		\theta' = \theta + \theta_{0}\left(R_{2} - r\right)/\left(R_{2}-R_{1}\right)~~~~\left(R_{1} < r < R_{2}\right),
	\end{align}
\end{subequations}
where $\theta_{0}$ denotes the rotation angle. With Eqs.~(\ref{E7}) and (\ref{E9}), we derive the rotator's thermal conductivity:
\begin{eqnarray}
	\overset{\leftrightarrow}\kappa_{2}' =
	\begin{pmatrix}
		\kappa_{rr} & \kappa_{rr}\frac{r'\theta_{0}}{R_{2}-R_{1}}\\ \kappa_{rr}\frac{r'\theta_{0}}{R_{2}-R_{1}} & \kappa_{rr}\left(\frac{r'\theta_{0}}{R_{2}-R_{1}}\right)^{2} + \kappa_{\theta\theta}
	\end{pmatrix},\label{E14}
\end{eqnarray}
which stands as the critical parameter for a thermal chameleon-like rotator, as long as $\tensor\kappa_{2}$ fulfills Eq.~(\ref{E3}).
\section{Simulations and experiments}
\begin{figure}
\centering
\includegraphics[width=1\linewidth]{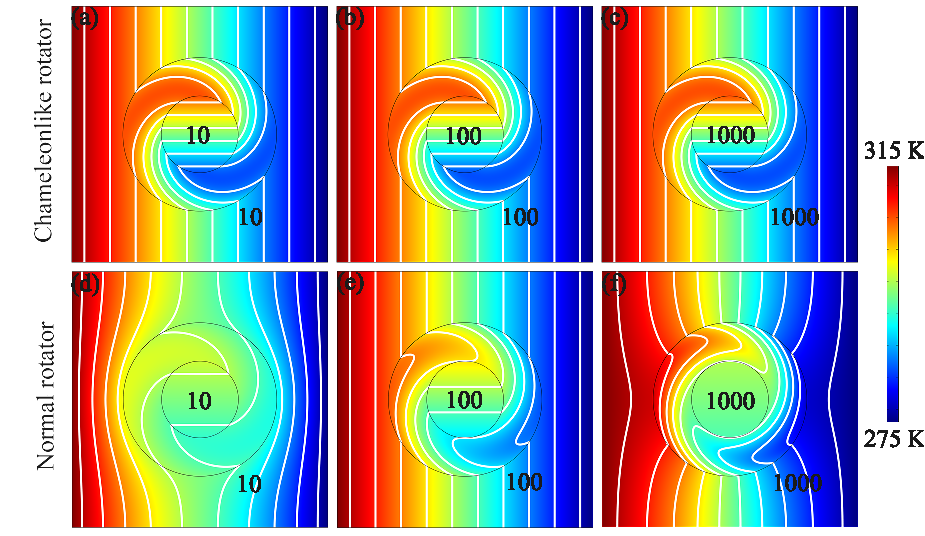}
\caption{\label{2} Simulation results of (a)-(c) chameleonlike rotator and (d)-(f) normal rotator. White lines represent isotherms, and the values in each simulation are the corresponding thermal conductivities. The system size is $1\times1$~m$^2$. The outer and inner diameters of the shell are 0.3 and 0.6~m, respectively. Adapted from Ref.~\cite{YangPRAP20}.}
\end{figure}

To validate the proposed approach, we initiated simulations using COMSOL MULTIPHYSICS (${\rm http://www.comsol.com/}$). The system under consideration corresponds to the one depicted in Fig.~\ref{1}(c). A comparison between a chameleon-like rotator and a conventional rotator is performed, as shown in Fig.~\ref{2}. Before undergoing rotation transformation, the thermal conductivities of the chameleon-like rotator and the normal rotator are ${\rm diag}\left(10^6,10^{-3}\right)$ and 100~W~m$^{-1}$~K$^{-1}$, respectively. Notably, the radial thermal conductivity of the transformation-invariant metamaterial should be significantly higher than the environmental thermal conductivity (at least two orders of magnitude) to ensure the chameleon-like rotator's functionality. Further investigation involves varying the environmental thermal conductivity from 10 to 1000~W~m$^{-1}$~K$^{-1}$, and the chameleon-like rotator consistently performs as expected—rotating heat flux without distorting the environmental temperature profile [see Figs.~\ref{2}(a)-\ref{2}(c)]. Consequently, simulation results affirm the chameleon-like property. However, the conventional rotator fails to exhibit this behavior. For an environmental thermal conductivity of 100~W~m$^{-1}$~K$^{-1}$, it behaves similarly to a traditional rotator [see Fig.~\ref{2}(e)]. When the environment changes, the temperature profile becomes distorted [see Figs.~\ref{2}(d) and \ref{2}(f)]. Thus, the normal rotator remains unresponsive to environmental changes.

\begin{figure}
\centering
\includegraphics[width=1\linewidth]{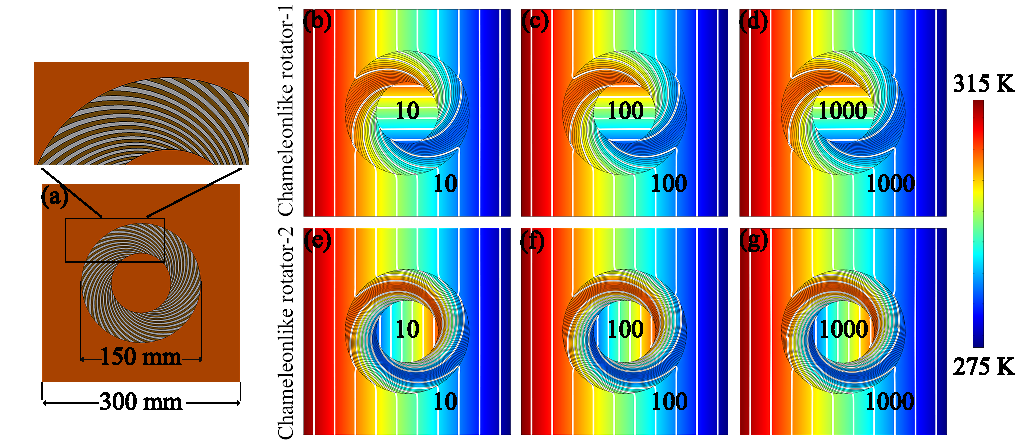}
\caption{\label{3} Simulation results of chameleonlike rotators with multilayered structures. (a) Schematic diagram. The structure is composed of two kinds of material with thermal conductivities of $10^6$ and $10^{-3}$~W~m$^{-1}$~K$^{-1}$, respectively. Simulations results of (b)-(d) chameleonlike rotator-1 and (e)-(g) chameleonlike rotator-2 in different environments. The composite materials in (b)-(g) are the same as those in (a). Adapted from Ref.~\cite{YangPRAP20}.}
\end{figure}

To experimentally validate our concept, finding a natural material that exactly satisfies Eq.~(\ref{E14}) proves challenging. Instead, we resort to the effective medium theory to realize the requisite parameter values. Drawing inspiration from multilayered structures~\cite{PRL2012}, we devise a chameleon-like rotator illustrated in Fig.~\ref{3}(a). In line with Eqs.~(\ref{E3}) and~(\ref{E14}), we opt for two materials possessing vastly contrasting thermal conductivities ($\kappa_l \approx 10^6$~W~m$^{-1}$~K$^{-1}$ and $\kappa_s \approx 10^{-3}$~W~m$^{-1}$~K$^{-1}$) to roughly fulfill Eq.~(\ref{E3}), and we incorporate a helical structure to achieve a semblance of Eq.~(\ref{E14}). Simulation outcomes are depicted in Figs.~\ref{3}(b)-\ref{3}(g). Specifically, Figs.~\ref{3}(b)-\ref{3}(d) reveal the results of chameleon-like rotator-1, rotating heat flux by 90 degrees. Figures~\ref{3}(e)-\ref{3}(g) illustrate the outcomes of chameleon-like rotator-2, accomplishing a 180-degree heat flux rotation. Hence, crafting chameleon-like rotators using multilayered composite structures is indeed feasible.

\begin{figure}
\centering
\includegraphics[width=1\linewidth]{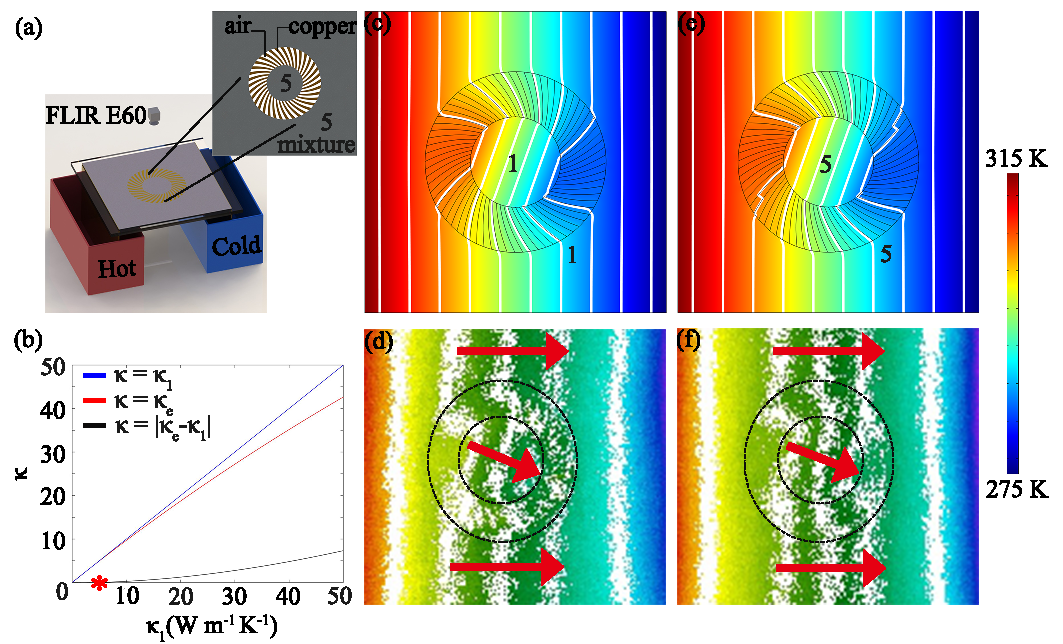}
\caption{\label{4} Laboratory experiments of chameleonlike rotator. (a) Experimental setup. The structure is composed of copper ($\kappa_{cu}\approx 400$~W~m$^{-1}$~K$^{-1}$) and air ($\kappa_{air}\approx 0.026$~W~m$^{-1}$~K$^{-1}$). (b) $\kappa$ as a function of $\kappa_{1}$. The blue (top) and red (middle) lines correspond to $\kappa_1$ and $\kappa_e$, respectively. The black (bottom) line refers to $|\kappa_{e} - \kappa_{1}|$. The coordinate of $*$ is $\left(5,\,0.047\right)$. (c) and (e) Simulation results and (d) and (f) experimental results of the samples. The arrows indicate the direction of heat flux. The inner and outer diameters of the shell are 0.075 and 0.15~m, respectively. Adapted from Ref.~\cite{YangPRAP20}.}
\end{figure}

Constrained by experimental conditions, we opt for copper ($\kappa_{cu}\approx$ 400~W~m$^{-1}$~K$^{-1}$) and air ($\kappa_{air}\approx$ 0.026~W~m$^{-1}$~K$^{-1}$) to construct a multilayered composite structure for a small-angle rotator. According to the series/parallel connection formula~\cite{JAP2017}, the effective thermal conductivity of the composite structure is roughly ${\rm diag}\left(200,0.052\right)$~W~m$^{-1}$~K$^{-1}$ prior to transformation. This imposes an upper limit on the range of environmental thermal conductivity variation. We calculate $\kappa_{e}$ while varying $\kappa_{1}$ from 0.1 to 50~W~m$^{-1}$~K$^{-1}$, confirming that the chameleon-like rotator operates effectively within the range of 0.1 to 5~W~m$^{-1}$~K$^{-1}$, as shown in Fig.~\ref{4}(b). The difference $|\kappa_{e} - \kappa_{1}|$ remains below 0.05~W~m$^{-1}$~K$^{-1}$ (marked by the star, with a deviation smaller than 1\%). Subsequently, we conduct experiments with environmental thermal conductivities of 1 and 5~W~m$^{-1}$~K$^{-1}$. The configuration of the system is depicted in Fig.~\ref{4}(a). The chameleon-like rotator comprises air and copper, fabricated using laser cutting. The environment consists of colloidal materials formed by mixing silica gel ($\kappa_{gel} = 0.15$~W~m$^{-1}$~K$^{-1}$ and density $\rho_{gel} = 1.14\times10^{3}$~Kg$~$m$^{-3}$) and white copper powder ($\kappa_{wcu} =33$~W~m$^{-1}$~K$^{-1}$ and $\rho_{wcu} = 8.65\times10^{3}$~Kg~m$^{-3}$). The thermal conductivity of the mixture is determined using the Bruggeman formula~\cite{APL2018}:
\begin{eqnarray}
	p_{gel}\frac{\kappa_{gel}-\kappa_{mix}}{\kappa_{gel}+2\kappa_{mix}} + (1 - p_{gel})\frac{\kappa_{wcu}-\kappa_{mix}}{\kappa_{wcu}+2\kappa_{mix}} = 0,\label{E15}
\end{eqnarray}
where $p_{gel}$ denotes the volume fraction of silica gel in the mixture. By setting $\kappa_{mix} = 1$ or $5$~W~m$^{-1}$~K$^{-1}$, we can derive the composition ratio of silica gel, which aids in fabricating the colloidal materials. Despite the existence of interface thermal conductance, the mixture of regions I and III possesses sufficient fluidity to ensure good contact between the object and copper. Subsequently, two tanks containing hot and cold water serve as the hot and cold sources, respectively. The temperature profile of the sample is measured using the FLIR E60 infrared camera. Experimental results are presented in Fig.~\ref{4}(d) ($\kappa_{mix} = 1$~W~m$^{-1}$~K$^{-1}$) and \ref{4}(f) ($\kappa_{mix} = 5$~W~m$^{-1}$~K$^{-1}$). Corresponding simulation outcomes are displayed in Figs.~\ref{4}(c) and \ref{4}(e). Despite the heat dissipation resulting from the connection of the sample to the hot and cold tanks via two copper plates, as well as the natural convection between the sample and air, there is a slight discrepancy between the computational and experimental values. Nevertheless, this discrepancy does not undermine the anticipated results. The isotherms remain unaltered despite changes in the environmental thermal conductivity. Furthermore, heat flux rotates as expected. Thus, the experimental results align well with the simulation results, confirming the viability of chameleon-like rotators.

\section{Discussion and Conclusion}
The fundamental distinction in our approach lies in commencing the coordinate transformation from a highly anisotropic parameter, which has been substantiated to exhibit chameleon-like behavior. Consequently, a rotator devised using this parameter inherits the chameleon-like attributes. Notably, this behavior remains invariant despite alterations in the rotator's shape, enabling the design of chameleon-like rotators with arbitrary geometries using our method. It's worth noting that an ideal transformation-invariant (highly anisotropic) shell adheres to Eq.~(\ref{E3}), underscoring that greater anisotropy yields enhanced chameleon-like behavior and an expanded operational range. Due to the scarcity of highly conductive materials, the fabricated rotator's operational range spans from 0.1 to 5~W~m$^{-1}$K$^{-1}$. 

In summation, we have introduced the concept of chameleon-like rotators harnessed through transformation-invariant metamaterials. Leveraging highly anisotropic thermal conductivity, these rotators can seamlessly operate across varied environments, streamlining efforts and resources. Both simulation and experimental results concur in validating the viability of this scheme. These findings enhance the intelligence of traditional thermal metamaterials, holding potential for applications in the realm of intelligent metamaterial design. Furthermore, this scheme's applicability can be extended to other domains, such as hydrodynamics, where the role analogous to thermal conductivity in thermotics is played by key parameters like permeability or viscosity.

\textit{Notes}. The main body of this conference paper refers to reference Ref.~\cite{YangPRAP20}.

\section*{Acknowledgments}
We acknowledge financial support by the National Natural Science Foundation of China under Grant No. 11725521, and by the Science and Technology Commission of Shanghai Municipality under Grant No. 20JC1414700.

\end{document}